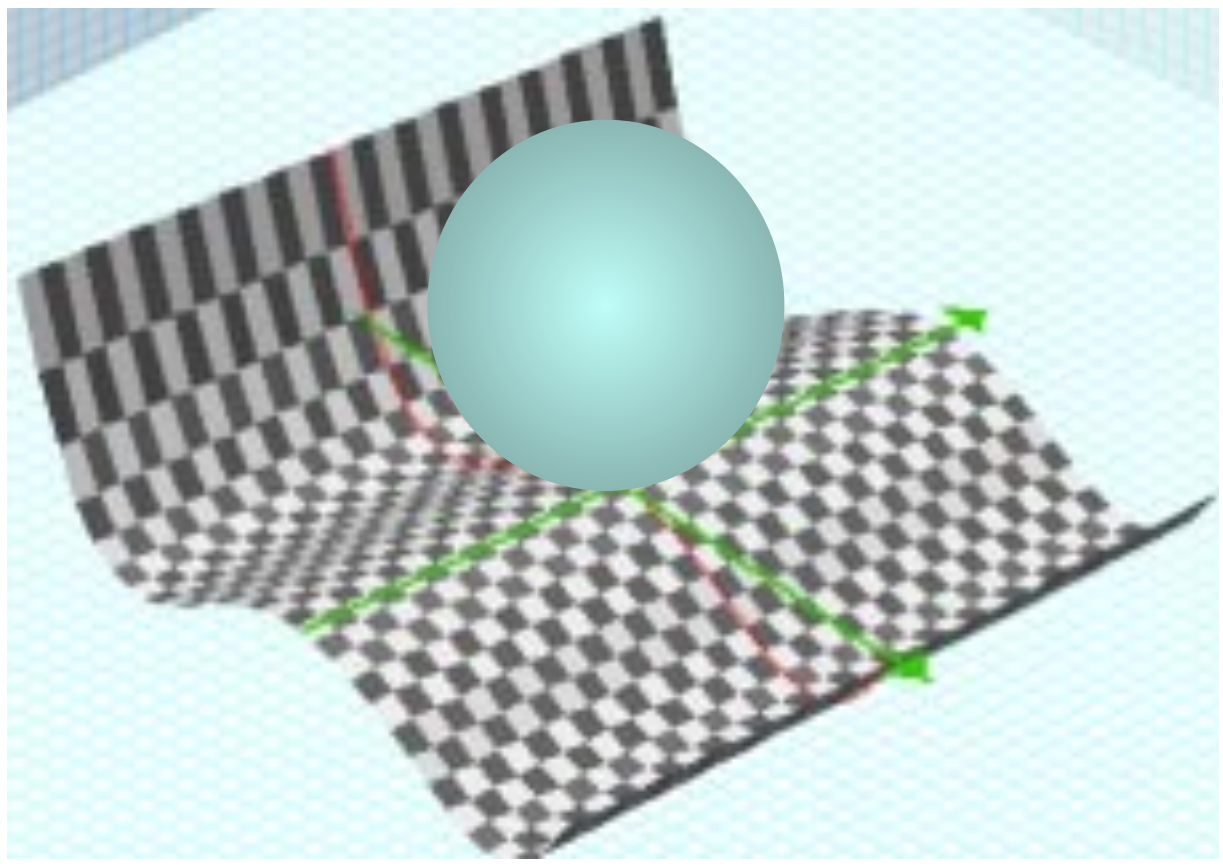

# Seeing the Forest in the Tree: Applying VRML to Mathematical Problems in Number Theory

Neil J. Gunther

Performance Dynamics Consulting™
Castro Valley, California 94552
njgunther@perfdynamics.com

# Seeing the forest in the tree: applying VRML to mathematical problems in Number Theory

Neil J. Gunther[†]

Performance Dynamics Consulting, Castro Valley, CA 94552

## ABSTRACT

Hamming claimed "the purpose of computing is insight, not numbers." In a variant of that aphorism, we show how the Virtual Reality Modeling Language (VRML) can provide powerful insight into the mathematical properties of numbers. The mathematical problem we consider is the relatively recent conjecture colloquially known as the "3x + 1 problem". It refers to an iterative integer function that also can be thought of as a digraph (or tree) rooted at unity with the other numbers in any iteration sequence located at seemingly randomized positions throughout the tree. The mathematical conjecture states that there is a unique cycle at unity. So far, a proof for this otherwise simple function has remained intractable. Many difficult problems in number theory, however, have been cracked with the aid of (often just 2-dimensional) geometrical representations. Here, we show that any arbitrary portion of the 3x+1 digraph can be constructed by iterative application of a unique subgraph called the G-cell generator—similar in concept to a fractal geometry generator. We describe the G-cell generator and present some examples of the VRML worlds developed programmatically with it. Perhaps surprisingly, this seems to be one of the few attempts to apply VRML to problems in number theory.

## 1. INTRODUCTION

This is not a mathematics paper. It's a paper about 3-dimensional imagery as a collaborative tool for exploring certain kinds of mathematical problems using the internet. In particular, we describe the application of the Virtual Reality Modeling Language (VRML) to the exploration of complexity inherent in a recent Number Theory conjecture. The conjecture remains mathematically unproven but it is our hope that the imaging methodology developed here might assist mathematicians seeking such a proof.

The mathematical problem that is the focus of this paper is a chestnut from the 1930's variously known as: *the 3x + 1 problem*, *the Collatz problem*, *Ulam's problem*, among other names. The problem can be stated quite simply. Start with any positive integer. If it is an even number, halve it. Otherwise, multiply the number by 3, add 1 to it, and then halve it (e.g., if the starting number was 7, the next integer in the sequence would be 11). Take the result and repeat the process. Inevitably, any such sequence seems to end up in the cycle 2, 1, 2, 1, etc., no matter the choice of starting number. By convention, one terminates the iterations at 1. References to the 3x + 1 problem continue to appear in unexpected places.[1,2,3.]

Several remarkable features are noteworthy about this innocent looking iterative function:

1) Being easy to program on a computer[2,3.], it has been shown to hold for starting integers exceeding $10^{13}$.
2) The length of any sequence bears no simple relationship to the magnitude of the starting number.
3) The values in any sequence appear to be completely erratic and unpredictable.

From the mathematician's standpoint, the first of these is of most interest because empirical evidence is not the same thing as a rigorous mathematical proof and $10^{13}$ is not a large number in mathematics. On the other hand, understanding inherent structure and symmetries within the sequences may be of value in ultimately establishing a proof for the conjecture. The latter tack is the one taken in this paper and it provides the motivation for applying VRML to the problem.

[†] Networking coordinates: mailto:njgunther@perfdynamics.com; http://www.perfdynamics.com/

From the engineering standpoint, number theory is important for understanding such applications as: cryptography,[4.] acoustics, and communications,[5.] to name a few. For which engineering systems the 3x + 1 function might be relevant is still an open question.

## 2. MATHEMATICAL MAPPINGS

The eminent mathematician Paul Erdos suggested "mathematics is not ready for this kind of problem."[6.] Whether his statement is correct or not, in all likelihood new techniques are being developed in an effort to prove the 3x + 1 conjecture. Many problems in number theory have been cracked with the aid of geometrical interpretations. The 3x+1 iterative function can be thought of as a directed tree rooted at 1 with the other numbers in any iteration sequence located at seemingly randomized positions in the tree. (See Fig. 2) The mathematical conjecture states that there is a unique cycle 2, 1, 2, 1, etc., in the tree. Although the function is stunningly simple, a rigorous proof has remained intractable. The reader who is interested in the deeper mathematics of the 3x + 1 problem is referred to the monograph by Wirsching[7.] which also contains a comprehensive bibliography.

Recently, the author has demonstrated that any arbitrary portion of the 3x + 1 tree can be generated by iterative application of a unique subgraph called the G-cell generator—similar in spirit to a fractal geometry generator.[8.] In this paper, we develop the G-cell as a VRML object. It's role in the 3x + 1 problem could be a significant step for several reasons:

- VRML permits the observer to zoom into structures at all scales (up to limitations of the computing platform).
- VRML provides the means to alter comparative visual perspective (cf. Tukey's data-spinning concept[9.]).
- VRML enables remote collaborators to convey their 3-dimensional ideas unambiguously on the web.

To set the stage for the G-cell representation and its later VRML expression, we first review the 3x + 1 mapping and its associated directed graphs in a little more detail.

### 2.1 Sequences

The 3x + 1 or Collatz map can be defined more formally as:

$$T(x) = \begin{cases} x/2 & \text{if x is even,} \\ (3x+1)/2 & \text{if x is odd.} \end{cases}$$

The original function described by L. Collatz used (3x + 1) for the odd transformation. Without incurring any loss of generality, it is more convenient for technical reasons to replace the odd transformation by (3x + 1)/2, as shown. Part of the fascination with this mapping arises from the generated sequences *appearing* random and unpredictable yet, the numbers are completely deterministic.

Consider the application of T(x) to the starting number $x_0 = 7$. Since 7 is odd we apply the (3x + 1)/2 component and find that the next iterate $x_1 = 11$. Since 11 is also odd we apply the (3x + 1)/2 component again and find $x_2 = 17$. The next iteration produces $x_3 = 26$, which is even. Division by 2 yields $x_4 = 13$, which is odd and so on until the value 1 is reached; at which point we terminate application of the T(x) function. The complete sequence is shown in Table 1.

**Table 1. The 3x + 1 sequence produced by $x_0 = 7$**

| **Step** | 0 | 1 | 2 | 3 | 4 | 5 | 6 | 7 | 8 | 9 | 10 | 11 |
|---|---|---|---|---|---|---|---|---|---|---|---|---|
| **Value** | 7 | 11 | 17 | 26 | 13 | 20 | 10 | 5 | 8 | 4 | 2 | 1 |

We see immediately how unpredictable the sequence looks. Moreover, it is not evident that the sequence should take 11 iterations to reach 1 and that the maximum value taken by any iterate is 26. The wildly varying nature of the iterates is displayed more vividly in Fig. 1 where the starting value is $x_0 = 27$ and 70 iterations are needed to reach 1. Further contrast this with a much larger starting number, $x_0$ = 1024. Since it is a power of two, the T(x) mapping reaches 1 by successive halving in just 10 steps i.e., fewer iterations than either 27 or 7. It is in this sense that the number of iterations is not proportional to the magnitude of the starting number.

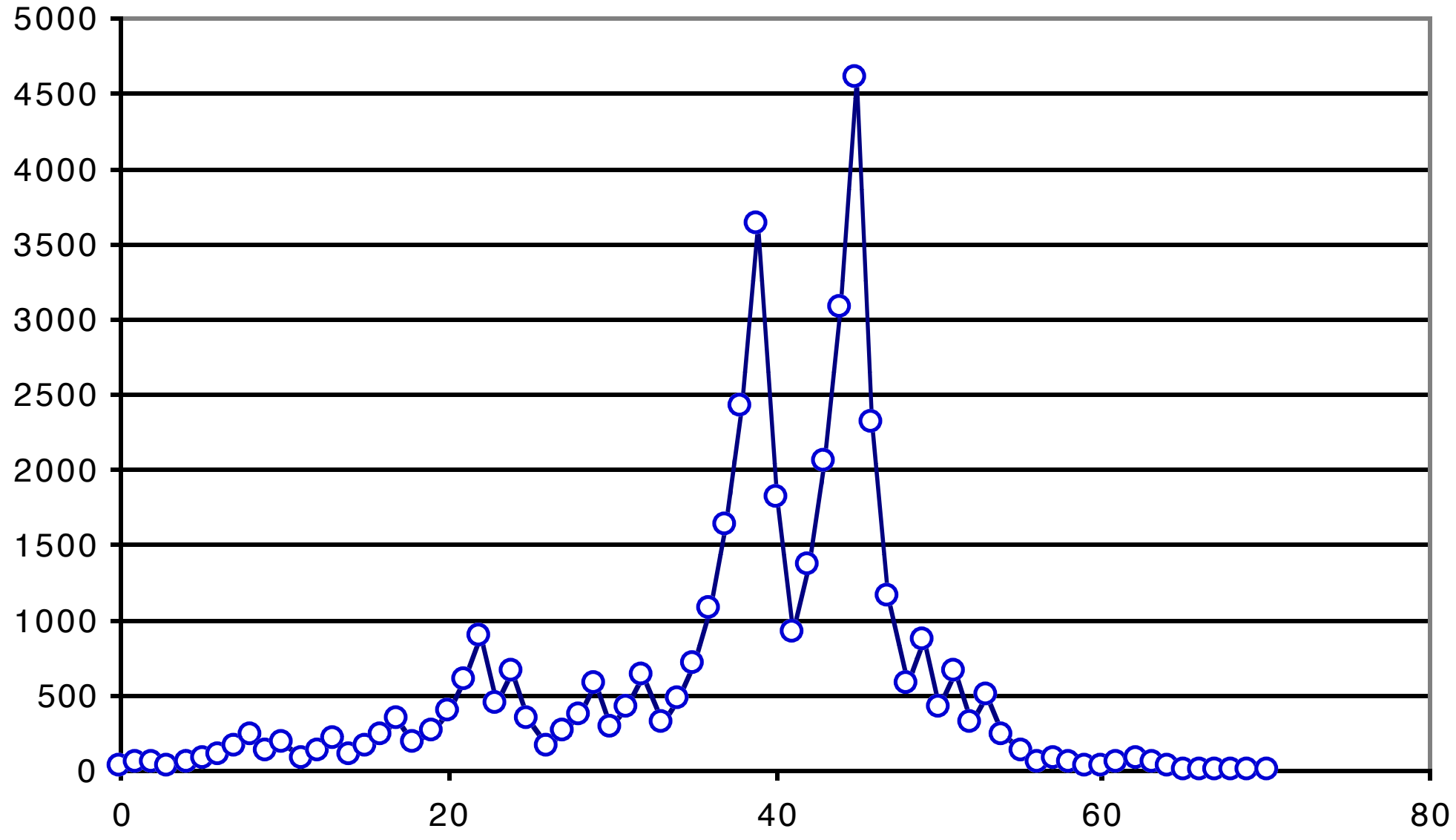

Figure 1. The erratic sequence produced by $x_0$ = 27 and the 70 iterations it takes to reach 1.

This erratic behavior of the 3x + 1 iterates can be better understood by arranging them into a directed graph.

**2.2 Trees**

An example of such a planar directed graph[10.] or tree is shown in Fig. 2.

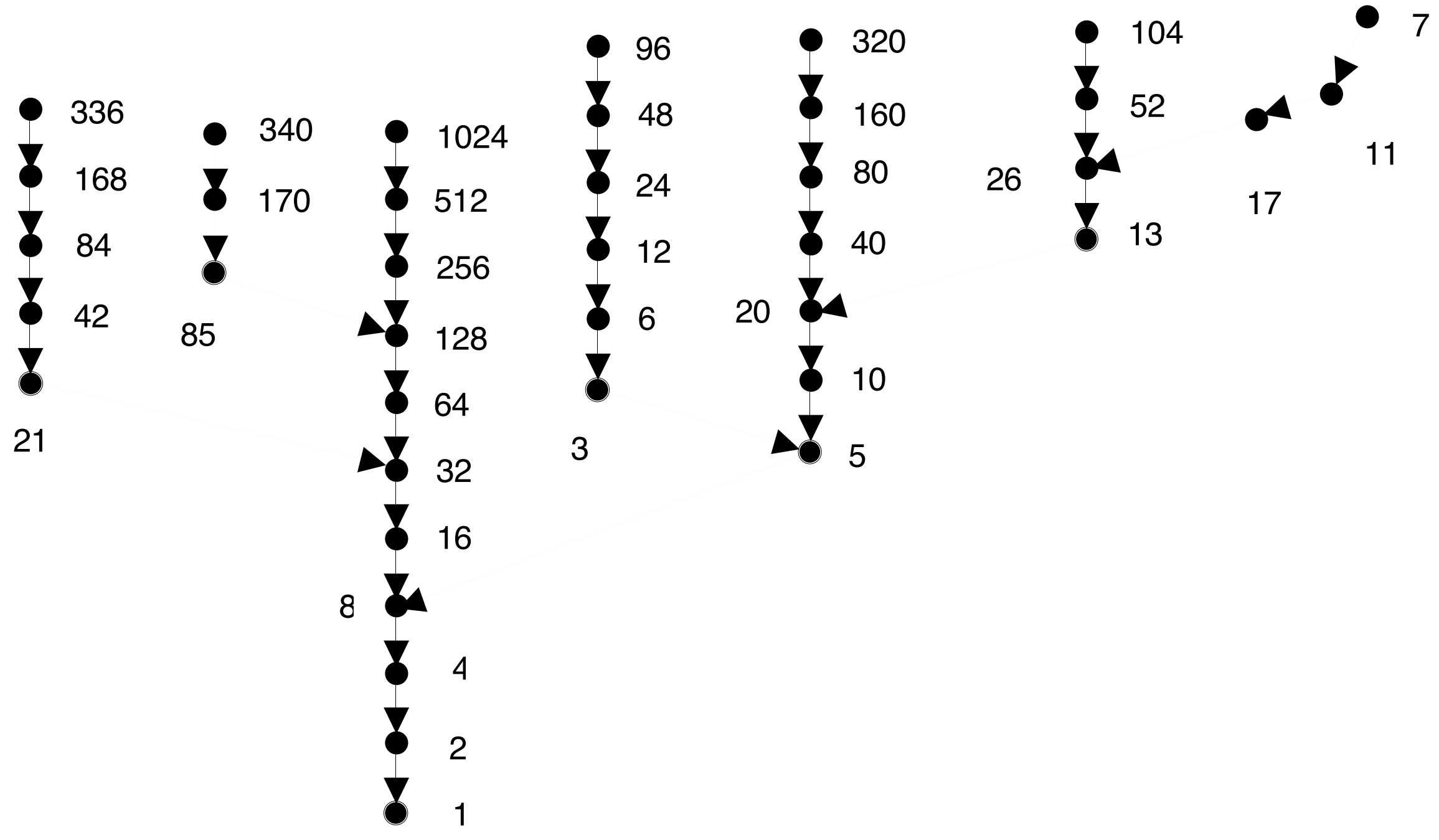

Figure 2. The 3x + 1 tree with some branches elided.

The tree is rooted at 1 near the center of Fig. 2 and all the numbers that lie vertically above 1 are powers of two. Other vertical branches commence at numbers like 3, 5, 13, 21, etc. Many of these vertical branches have branchings higher up and others (such as the branch that starts at 3) have no branchings at all. These branchings correspond to the odd-even discrimination defined by T(x).

The sequence in Table 1 can now be understood as being analogous to the motion of ball-bearing falling under the influence of gravity from node to node in the direction of each arc. The even map corresponds to vertical decent while the odd map corresponds to sideways motion toward to the main power-of-two trunk. Although the falling motion is uniform, the numbers through which the ball-bearing passes appear random because of the way the nodes in the tree are decorated.

It is amusing to note that it was attempts to understand the mathematical properties of certain integer functions as graphs that led L. Collatz (circa 1930) to consider the T(x) mapping in the first place.[7.]

## 3. G-CELL GENERATOR

The remainder of this paper relies on the observation that the tree in Fig. 2 can be constructed from a unique generating element, hereafter referred to as the *G-cell*. In Section 4, we show how this unique generator can be combined with the VRML language to develop immersive 3-dimensional structures for mathematical exploration.

### 3.1 G-cell Construction

The G-cell is a directed acyclic graph[10.] (DAG) comprising seven nodes and six arcs. For our purposes, we wish to arrange these nodes and arcs in such a way that we can layout the 3x+1 tree in a highly regular geometrical fashion. To achieve this, we associate the nodes of the G-cell with the vertices of a grid indicated by the dashed lines in Fig. 3a. Because there are only seven DAG nodes, one is absent from the grid point midway between the lower nodes labeled A and B. In addition, since there are only six arcs, no arc is present on the grid edge between the nodes labeled 4B and 4B + 1. The G-cell arrangement lies on four unit-cells of the grid.

Since the G-cell is an acyclic graph, it is not closed like an 'O' but open like a 'C' or, more accurately, the asymmetric letter 'G.' Hence, the name. In Fig. 3a, the 'G' shape has been swiveled by 180 degrees about a imaginary diagonal passing through nodes A and 4B. This particular orientation is arbitrary and simply corresponds to an organization of the 3x + 1 tree with the root in the lower part of the page; similar to that shown in Fig. 2. We adhere to this orientation of the G-cell for the remainder of this section.

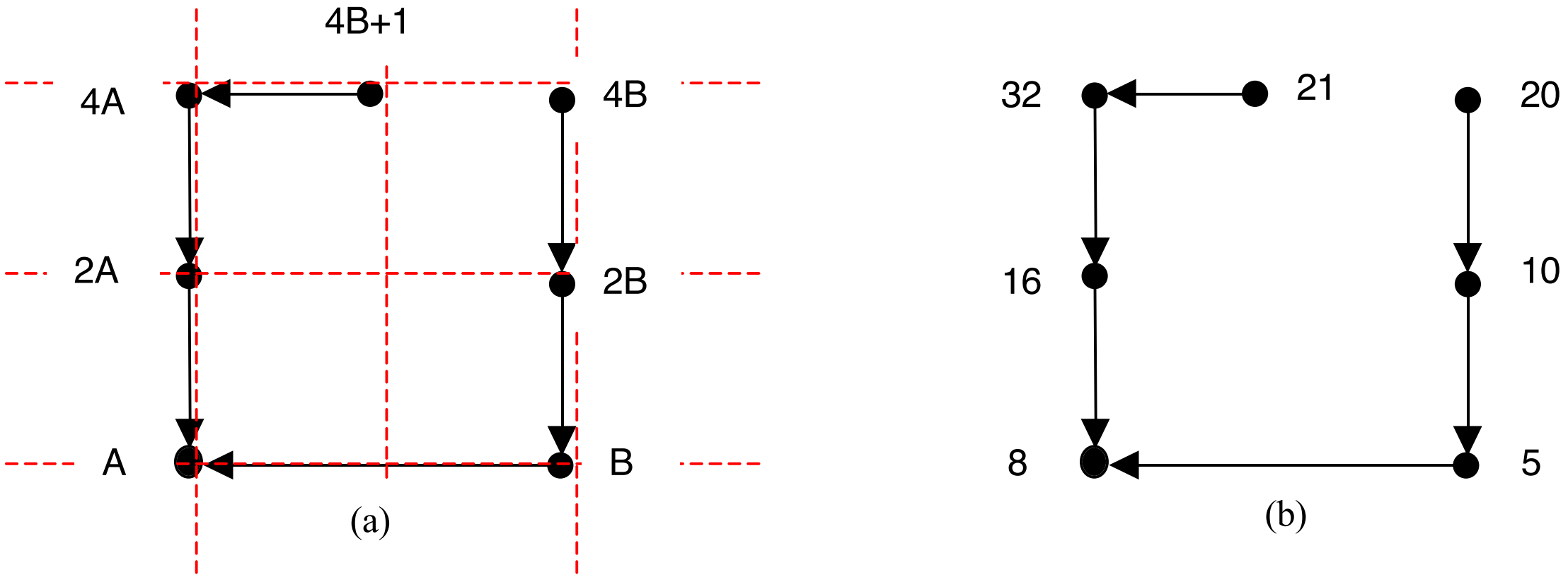

Figure 3. Algorithm for calculating the value of G-cell nodes.

As in Fig. 2, the nodes of the G-cell take on integer values. The value of a G-node is determined according to the following algorithm. Start at the lower left node labeled A in Fig. 3a. The node above it has twice the value (i.e., 2A) and the node above that has twice its value (i.e., 4A). A numerical example is given in Fig. 3b with A = 8, 2A = 16, and 4A = 32.

The lower right node, B, has a value given by, 3B = (2A – 1), and is joined by an arc to node A. In the numerical example, node B = 15/3 = 5. We note in passing that since all nodal values must be integral, node B can only exist if the following condition holds: (2A – 1) mod 3 = 0. In other words, if the value, 2A – 1, is not divisible by 3, node B cannot exist. For example, if we had started with the value A = 3, then 2A = 6, and since 5 mod 3 ≠ 0, node B is not permitted. This simply means that there cannot be any extension of the G-cell to the right of node A. We elaborate on this point in the next section when we discuss how G-cells can be connected together.

Continuing with our example, the right-hand side of the G-cell is developed by successive doubling (i.e., B, 2B, 4B) in a manner identical to that used for the left-hand side. So far, this algorithm only produces six of the nodes and five of the arcs. The seventh node is located between nodes 4A and 4B but is connected only by an arc to node 4A. Its value is 4B + 1 (e.g., the seventh node in Fig. 3b has the value 4B + 1 = 20 + 1 = 21).

How does the G-cell relate to 3x + 1 function? Successive iterations of the function T(x) correspond to traversing neighboring node pairs along an arc. Referring to Fig. 3b, we see that starting at node 20 (an even number) there is only one vertical arc to traverse and it connects to node 10 and then by another arc to node 5. Since 5 is an odd number, we now move along a left-pointing horizontal arc to node 8 and so on down to node 1. For this orientation of the G-cell (corresponding to node 1 residing in the lower left of the tree) the arcs of the G-cell point either down or to the left.

Two other features of the G-cell are noteworthy. Firstly, the sides of the G-cell correspond to a T(x) subsequence. The subsequence just described in Fig. 3b is also a subsequence of the one defined in Table 1. Secondly, although the subsequence of arcs starting at node 21 in Fig. 3b are not traversed by the sequence commencing at node 7 (see Fig. 2), they would be traversed by possible other sequences e.g., those starting at node 42 or node 84.

The virtue of the G-cell for our purposes is that neighboring nodes can be constructed around any starting node or any subsequence. The relationship *between* subsequences could be just as important as any individual sequence. Making such clusters of nodes and subsequences manifest requires being able to combine multiple G-cells correctly in some arbitrary neighborhood, and that is the subject of the next section.

### 3.2 Connecting G-cells

In this section, we treat the G-cell as a subgraph. As noted in Section 3.1, the existence of the lower right node (labeled B in Fig. 3a), is determined by the condition (2A – 1) mod 3 = 0. The extension of this rule gives a set of rules which determine how G-cells abut one another to generate an arbitrary portion of the 3x + 1 tree.

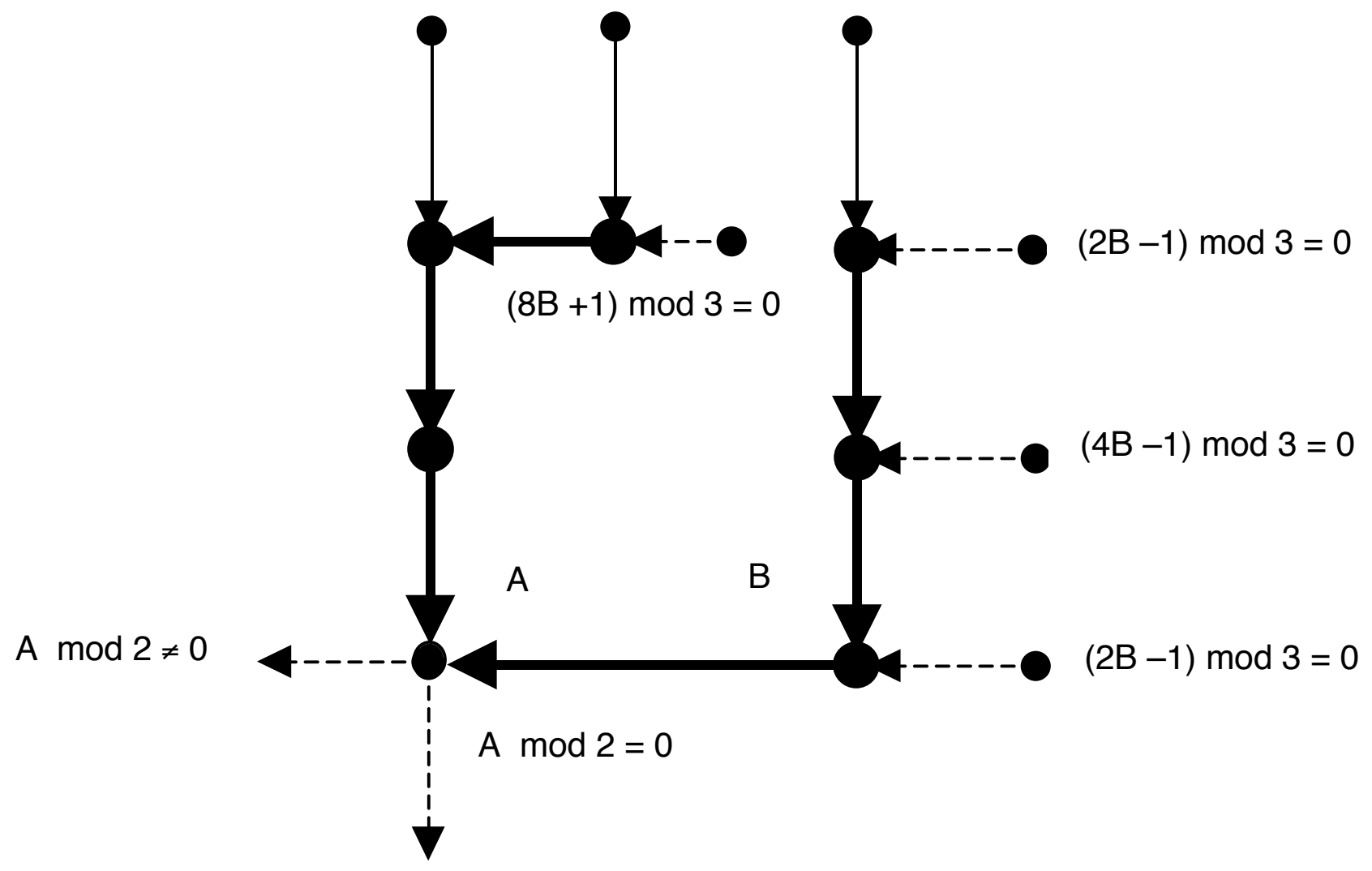

Figure 4a. Calculating G-cell connections.

There are only two types of connections possible: vertical (corresponding to doubling or halving of node values) and horizontal (generated by odd-valued nodes). Using this convention, the rules governing the existence of connections between G-cells are summarized in Fig. 4a where, for reference, the G-cell is shown with bold arcs and nodes. Since all three nodes on the top row of the Gcell can be extended indefinitely upward by doubling, they are depicted as solid arrows. In fact, the two nodes on the left-hand side of the top row (denoted by 4A and 4B + 1 in Fig 3a), form the base of the next generation G-cells.

Continuing on the left side of Fig. 4a, there will be another G-cell of the same generation to the left of the current G-cell (shown in bold), if and only if node A is an *odd* integer i.e., A mod 2 ≠ 0. Otherwise, node A is *even* and the only possible connection is downward. Since either type of connection is permitted, they are shown as dashed arrows. Moreover, if there is no G-cell abutting the left side of the bold G-cell, then there cannot be any horizontal connections at node 2A or node 4A. Similarly, if a G-cell does abut the left-hand side of the current G-cell, there cannot be any horizontal connections at nodes 2A or 4A because a G-cell has no internal connections.

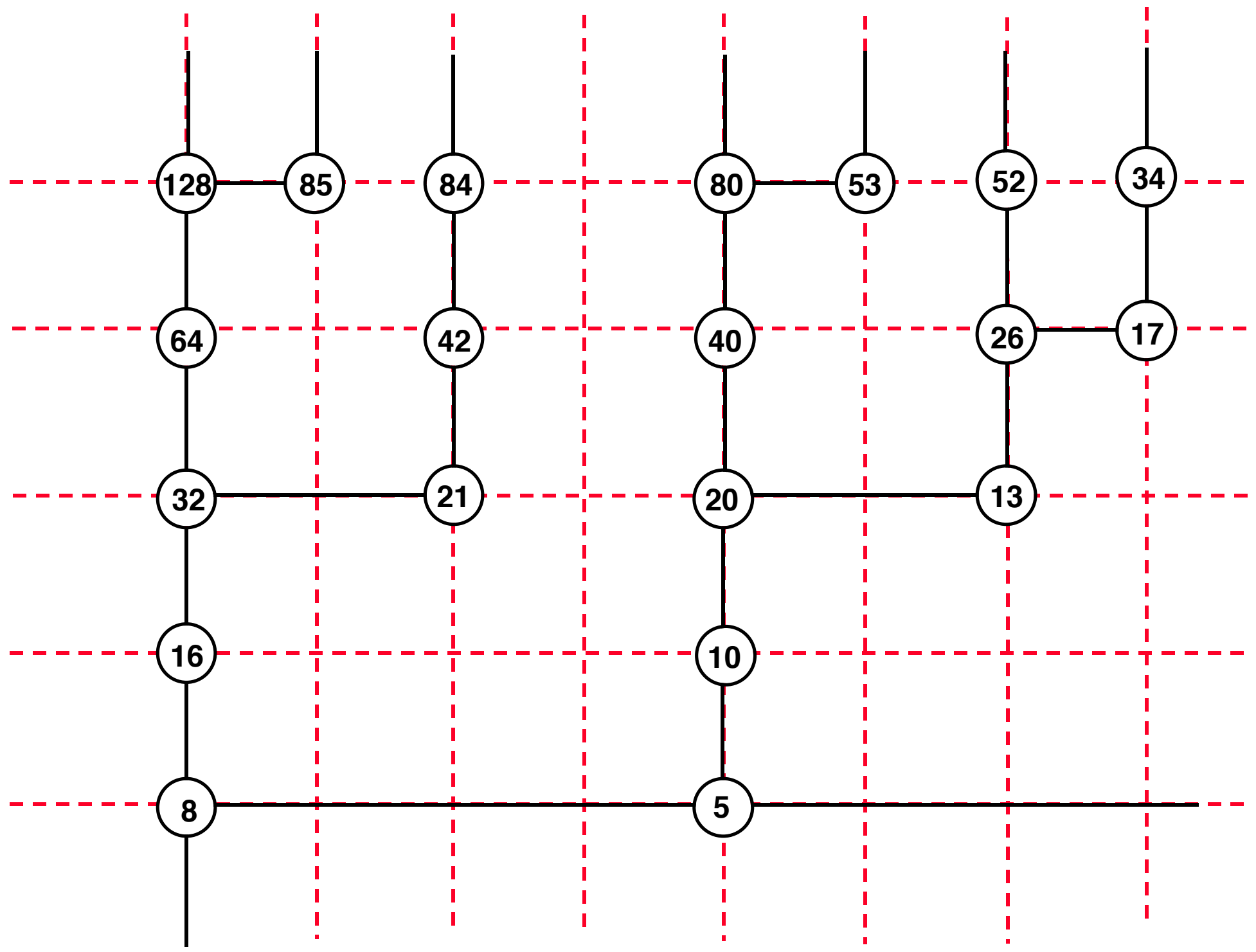

Figure 4b. Example of G-cells connected on the unit cell grid.

Referring now to the right-hand side of the bold G-cell, there cannot be any downward connection at node B since it is odd (in order that the current G-cell be defined at all). Whether there is a G-cell to the right with its bottom row of nodes abutting at node B or node 2B is determined as follows. If the condition (2B – 1) mod 3 = 0 holds, then there will be a G-cell of the same generation connected to the current G-cell at nodes B and 4B. Otherwise, the abutting G-cell will be connected at nodes 2B and 8B (not shown in Fig. 4a). A more complete set of connected G-cells is shown in Fig. 4b.

The overriding exception occurs when node B = 3(2k + 1) for any k = 0, 1, 2, …. For example, if k = 0 then B = 3, and all nodes above node 3 are determined by simple doubling i.e., B = 3, 2B = 6, 4B = 12, etc. Each of these nodes is already divisible by 3. Subtracting 1 from any of these nodes means that (2B – 1) mod 3 ≠ 0 and (4B – 1) mod 3 ≠ 0. So, there cannot be any horizontal branching off the vertical branch of node-pairs that starts with node 3. Similarly for other vertical branches that begin with a value given by the same k–sequence. All of these statements can be defined with greater rigor, but this is not a paper on the mathematics of 3x + 1.

The common terminating sequence 8, 4, 2 ,1 is represented by a G-cell which has degenerated into a digraph with a cycle between nodes A = 2 and B = 1 as depicted in Fig. 5. The right-hand side of the G-cell, above node 1, has now become a phantom subgraph. Another formulation of the 3x + 1 conjecture, therefore, would be: the 3x + 1 tree is comprised entirely of a network of G-cells that are connected (but open) except for the G-cell containing node 1; which must have a cycle under the requirement that all nodes be unique.

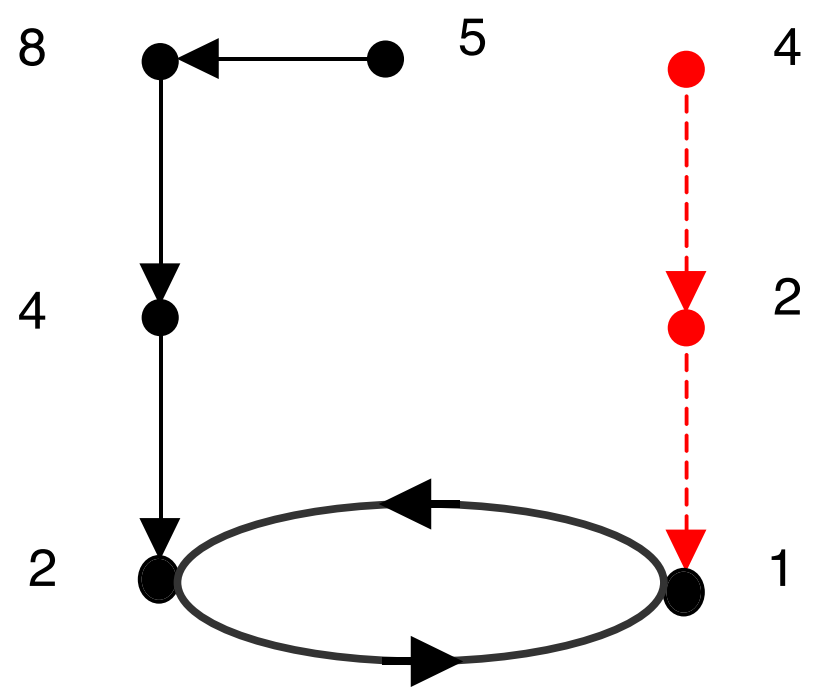

Figure 5. Root G-cell with cycle and the phantom node-pair 4-2.

For the purpose of this paper, the advantage of G-cell generator is that it can be used to render arbitrary portions of the 3x + 1 graph programmatically by abutting G-cells using the above rules. Figures 7 and 8 were constructed in this manner using the Perl[11.] programming language to emit VRML.

## 4. VRML REPRESENTATIONS

This author was introduced to the 3x + 1 problem in 1989, but it was not until 1993 that the concept of the G-cell first occurred to him. That led to an implementation of the 3x + 1 tree using a child's Tinker Toy® construction set. One quickly realizes, however, the severe limitations of this approach. Apart from exhausting the supply of parts very quickly, the structure becomes mechanically unstable and requires an additional supporting framework. Moreover, there are only a relatively small number of parts available and having constructed one portion of the tree, it must be torn down to construct another.

Clearly, rendering the G-cell graph on a computer with a mechanical or molecular CAD system would have been a better choice. But no such system was available to the author at the time, and the level of interest in 3x + 1 being no more than a hobby, the G-cell concept was never developed. In the meantime, the web arrived with browsers and more importantly VRML plug-ins. Spurred on by growing interest[12.] in the 3x + 1 problem, the author recently rendered the G-cell in VRML.[13.] This section describes that effort.

### 4.1 What is VRML?

VRML is an ISO standard programming language intended for the construction and distribution of interactive three dimensional content on the web. Standardization is the responsibility of the Web 3-D Consortium.[14.] In addition, the MPEG-4 specification will adopt VRML for its 3-D component. VRML is interoperable with other programming languages and uses JavaScript for its internal scripting, while Java is used as the external authoring interface.

VRML code is flat ASCII text and is intended to be read and edited by humans. VRML can be created by a dedicated software package designed specifically for building VRML worlds, exported from a 3-D modeling viewer (often a browser plug-in), or it can be created manually using a text editor. By current convention, VRML files have a .wrl extension. The sample VRML code in Fig. 6, shows how VRML builds compound 3-dimensional objects from simpler 3-dimensional primitives such spheres and cylinders. In this sense, VRML representations are already three dimensional.

```
Group {
    children [
      Shape {
          geometry
            Sphere {
                  radius 0.25
            }
      }
    ]
}
```

Figure 6. Sample VRML code representing the spherical nodes in Figures 7 and 8.

Although VRML objects are defined in the language as static objects with respect to a screen-based coordinate system, they are rendered dynamically. In other words, once a compound object is rendered initially in the VRML browser using the definitions contained in the text-based description, the user is free to move in 3-dimensions with respect to that object e.g., move inside the object, view the object from any angle, and interact with it in real-time. The burdens of maintaining visual perspective and cropping are handled in the VRML browser as if by magic!

### 4.2 G-cell Networks in VRML

Based on the G-cell concepts presented in section 3, we used the Perl programming language[11] to construct a portion (i.e., an 11 x 11 grid) of the 3x + 1 tree as a VRML world for integers less than 1024; approximately 32 integers. The results provide a proof of concept for exploring this mathematical problem with VRML tools.

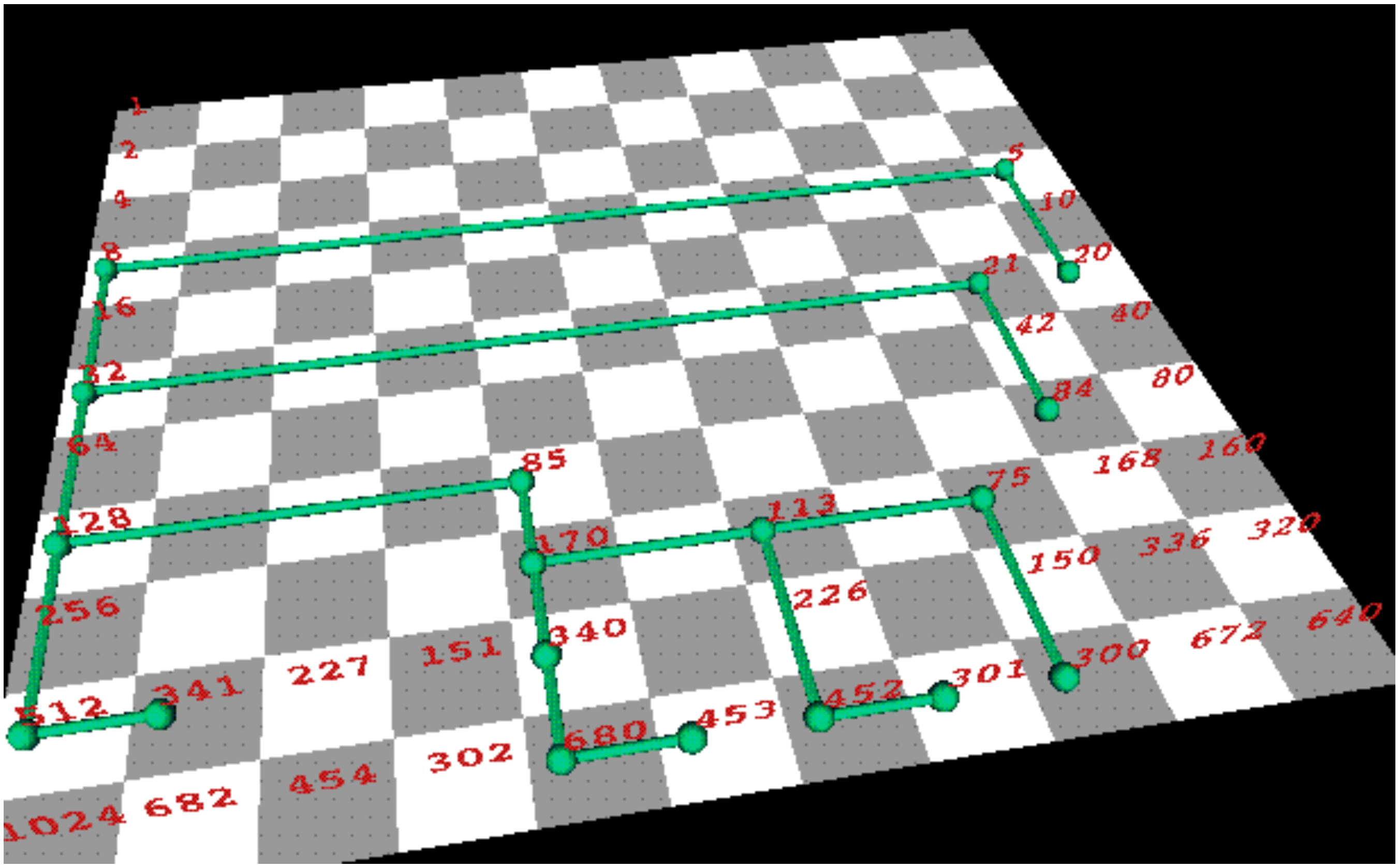

Figure 7. The G-cell construction of the 3x + 1 tree in VRML for some of the integers smaller than 1024.

One view of our results uses the elevation grid shown in Fig. 7. This has the appearance of looking down on a chess board and provides a simple organization of the visual field. Numbers sit above the VRML elevation grid arranged according to their G-cell nodal values using the algorithms of section 3. Interposed between the elevation grid and the array of numbers is the network of arcs and nodes belonging to the G-cell representation of the 3x + 1 tree.

In this view, the 3x + 1 tree is rooted at the top-left of the frame in accordance with the usual computer science convention. Scanning left to right in Fig. 7, the ‘G’ shape of the connected cells is quite apparent. Only complete and connected G-cells are shown. The root G-cell and other partial G-cells have been elided for clarity. Fig. 7 shows the 3x + 1 sequences represented as a formal tree in the mathematical sense (i.e., a planar digraph). Planarity requires that the G-cells be of different sizes in each generation; larger for small integers and smaller for large integers.

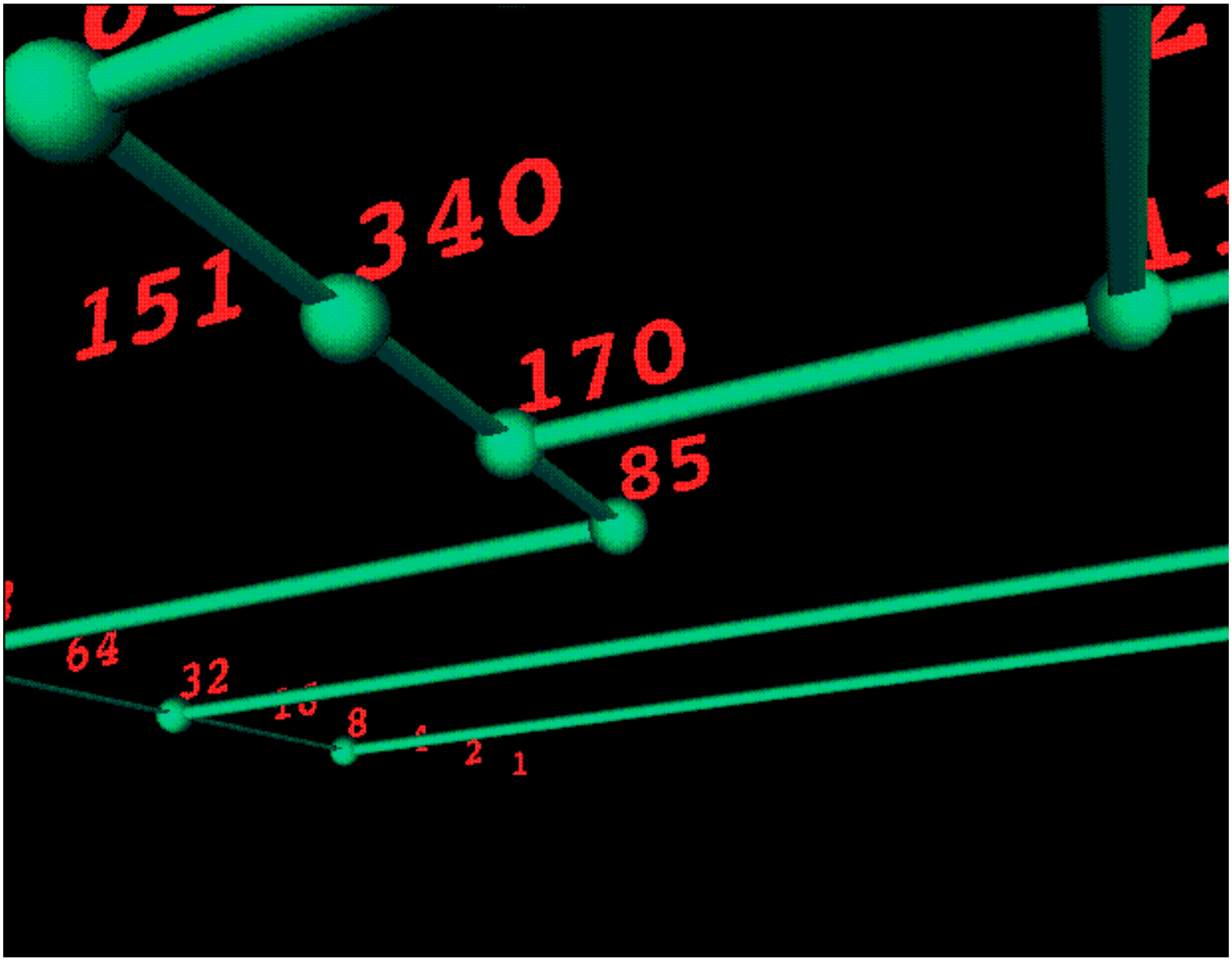

Figure 8. The 3x + 1 tree as seen by “flying” under the VRML G-cell network.

Figure 7 also shows the possible G-cell abutments. The power-of-two branch appears on the left hand side (cf. Fig. 4). The largest G-cell abuts the power-of-two branch at nodes 8 and 32 with a gap between nodes 20 and 21. The next largest G-cell abuts the power-of-two branch at nodes 32 and 128 with a gap between nodes 84 and 85. It, in turn, abuts the largest G-cell at nodes 21 and 32. The next generation G-cell abuts the power-of-two branch at nodes 128 and 512 with a gap between nodes 340 and 341. The next generation of G-cells that would otherwise fill that gap have been elided for clarity. But a similar string of G-cells abuts this one at nodes 170 and 680 with a gap between nodes 452 and 453 and the next G-cell abuts this one and has a gap between nodes 300 and 301. A complete network for numbers less than 1024 extends for four chessboard widths to the right that main power-of-two branch and is elided for clarity.

Another powerful aspect of the VRML representation (and one that cannot be conveyed by means of this static, 2-dimensional, paper) is the ability of the observer to move, zoom, rotate, and slide the G-cell construction. Figure 8 is a very meager attempt to give the reader some idea of how dramatic this effect can be. It is intended to convey the fractal-like quality of the G-cell generator. The 3x + 1 tree, produced in this way, allows the observer to 'travel' around or zoom into and out of the structure over a broad range of scales in much the same way as has become commonplace in the rendering of fractals. The interested reader, armed with a VRML browser[15], is invited to visit the author's website[13.] and personally experience these effects.

Our VRML representation does not demand that the 3x + 1 graph be rendered in the plane. We have presented it that way here to validate the concept of the G-cell as a feasible means to generate the 3x + 1 tree. But VRML also enables a volume representation involving multiple planes where nodes, that would otherwise overlap in the plane, are offset into a third dimension. To the mathematician, such a multidimensional representation would seem to be unnecessary on the face of it because 3x + 1 can already be represented as a planar graph like Figure 2. If, however, there is some latent dimensional dependencies in the problem, VRML could be useful for revealing it.

## 5. CONCLUSIONS

The G-cell defined as a VRML object enables new visualizations of the 3x + 1 problem in number theory. Even the limited VRML construction described in this paper permits the observer to zoom into structures over a broad range of scales, up to limitations of the computing platform. It also provides alternative 3-dimensional visual perspectives that are otherwise difficult to achieve and it enables remote collaboration over the internet. We remain cautiously optimistic about its use because, as Wirsching[7.] warns, many alternative representations of the problem have been tried but they appear to be as intractable as the 3x + 1 problem itself.

Throughout this paper we have focused on a tree representation of the 3x + 1 problem thereby making contact with the typical approach taken by mathematicians. Unlike standard graph theory, however, we imposed an organizing geometry onto the set of graph nodes and arcs belonging to the G-cell. This rectangular structure permits us to programmatically generate arbitrary portions of the 3x + 1 tree by iterating upon the G-cell generator. The concept is similar to a fractal generator but differs in that it is not recursively self-similar in the usual fractal sense. As far as this author is aware, it has not been previously recognized that a construction of the 3x + 1 digraph could be based on iterative application of a single geometrical primitive.

The G-cell construction in VRML also enables easier comparison across multiple sequences (or paths) simultaneously. Neighboring sequences may be as important as neighboring nodes in a single sequence. The multidimensional capabilities of VRML are potentially very useful for uncovering any latent dimensionality in the 3x + 1 problem.

A further enhancement to the current implementation would facilitate the observer to specify interactively any finite region of interest in the digraph and then programmatically generating the VRML graph according to those specifications.

## ACKNOWLEDGEMENTS

The author wishes to thank G. B. Beretta for several stimulating discussions (verbal and electronic) and for his suggestion to render the G-cell in VRML. I am also grateful to G. J. Wirsching for making his monograph freely available to me.